\documentclass[aip,jcp,amsmath,amssymb,reprint]{revtex4-1}

\usepackage{graphicx}% Include figure files
\usepackage{bm}% bold math

\usepackage[utf8]{inputenc}
\usepackage[T1]{fontenc}
\usepackage{mathptmx}

\usepackage[pdfusetitle,colorlinks,linkcolor=blue,citecolor=blue,urlcolor=blue]{hyperref}

\def\equationautorefname~#1\null{Eq.~(#1)\null}

\newcommand{\expect}[1]{\langle{}#1\rangle}
\newcommand{\corr}[1]{\expect{#1}_C}

\allowdisplaybreaks{}

\begin{document}

\title[Cumulant expansion for light-matter interaction]{Cumulant expansion for the treatment of light-matter interactions in arbitrary material structures}

\author{M. Sánchez-Barquilla}
\author{R. E. F. Silva}
\author{J. Feist}
\email{johannes.feist@uam.es}
\affiliation{Departamento de Física Teórica de la Materia Condensada and Condensed Matter Physics Center (IFIMAC), Universidad Autónoma de Madrid, E-28049 Madrid, Spain}

\date{\today}

\begin{abstract}
Strong coupling of quantum emitters with confined electromagnetic modes of
nanophotonic structures may be used to change optical, chemical and transport
properties of materials, with significant theoretical effort invested towards a
better understanding of this phenomenon. However, a full theoretical description
of both matter and light is an extremely challenging task. Typical theoretical
approaches simplify the description of the photonic environment by describing it
as a single or few modes. While this approximation is accurate in some cases, it
breaks down strongly in complex environments, such as within plasmonic
nanocavities, and the electromagnetic environment must be fully taken into
account. This requires the quantum description of a continuum of bosonic modes,
a problem that is computationally hard. We here investigate a compromise where
the quantum character of light is taken into account at modest computational
cost. To do so, we focus on a quantum emitter that interacts with an arbitrary
photonic spectral density and employ the cumulant or cluster expansion method to
the Heisenberg equations of motion up to first, second and third order. We
benchmark the method by comparing with exact solutions for specific situations
and show that it can accurately represent dynamics for many parameter ranges.
\end{abstract}

\maketitle

Light-matter interaction is of paramount importance for unraveling the laws of
nature and its deep understanding allows us to control and manipulate physical
and chemical systems. In particular, one can modify the properties of a quantum
emitter simply by changing its electromagnetic environment, for example by
enclosing it within an optical cavity. This may give rise to a change of the
decay rate for spontaneous emission in the weak coupling regime, the so-called
Purcell effect~\cite{Purcell1946}, or to the appearance of hybrid light-matter
states, so-called polaritons, in the strong-coupling regime~\cite{Kaluzny1983,
Thompson1992,Weisbuch1992,Lidzey1998}. Over the last decades, it has been shown
that strong light-matter coupling can be achieved using a large variety of
physical implementations as the ``cavity'' that provides the electromagnetic
field confinement. These include Fabry-Perot cavities consisting of two
mirrors~\cite{Lidzey1998}, propagating surface plasmon
polaritons~\cite{Bellessa2004}, plasmonic hole~\cite{Dintinger2005} and
nanoparticle arrays~\cite{Rodriguez2013}, isolated plasmonic
nanoparticles~\cite{Zengin2015} and nanoparticle-on-mirror
geometries~\cite{Chikkaraddy2016,Li2016Transformation}, as well as hybrid
cavities combining plasmonic and dielectric
materials~\cite{Gurlek2018,Franke2019,Bisht2019}. In many of these systems, the
electromagnetic field modes are not well-described by isolated lossy cavity
modes, and a correct treatment demands theoretical approaches that are able to
deal with the complexity of the electromagnetic field modes and their spectrum.

In principle, to treat the problem of light-matter interaction, one can rely on
the most general theory that describes light and matter on equal footing, i.e.,
quantum electrodynamics (QED)~\cite{Cohen-Tannoudji1987}. However, treating all
light and matter degrees of freedom in the systems described above in a quantum
mechanical way is an intractable problem and approximations must be performed.
One of the most common assumptions in quantum optics is to consider that
material system of interest only interacts with a single mode of the
electromagnetic (EM) field, with the interaction typically treated within the
dipole approximation. This leads to the Rabi~\cite{Rabi1937},
Dicke~\cite{Dicke1954}, Jaynes-Cummings~\cite{Jaynes1963} and
Tavis-Cummings~\cite{Tavis1968} models depending on the number of treated
emitters and the approximations performed~\cite{Carmichael1999,Gardiner2004,
Grynberg2010,Garraway2011,FriskKockum2019}, all of which have been successfully
used and extended to describe a wide variety of experimental implementations.
Nevertheless, as discussed above, the simplification to a single (or few)
quantized light modes in the treatment of the electromagnetic field is not
always a good approximation.

In some cases, the quantum character of the electromagnetic field may be
neglected and it is possible to rely on Maxwell's equations. In such mean-field
approaches, the classical EM field is then coupled to the dipole of
the quantum emitters and the coupled Maxwell-Schrödinger or Maxwell-Bloch
equations are solved~\cite{Pusch2012,Sukharev2017}. This approach in principle
allows for the description of arbitrary photonic structures, but misses all
effects due to the quantization of the EM field, such as spontaneous emission.
Recently, several groups have extended these approaches to allow for a more
complete description, based on, e.g., an Ehrenfest+Relaxation
approach~\cite{Chen2019I,Chen2019II} or cavity quantum electrodynamics with
multi-trajectory Ehrenfest dynamics~\cite{Hoffmann2019Capturing}.

In cases that a full quantum description is desired, a strategy has to be used
to quantize the EM field modes in the presence of material bodies. This is
possible for simple geometries using a variety of
strategies~\cite{Trugler2008,Waks2010,Li2016Transformation}. For systems with a
few, but possibly interfering, resonances, it was recently shown how to quantize
the corresponding quasi-normal modes as lossy cavity modes~\cite{Franke2019}.
For arbitrary material structures, the most general solution is given by the
framework of macroscopic QED~\cite{Huttner1992,Scheel1998,Scheel2008,
Buhmann2012I}, which was developed in the last few decades to circumvent the
problems that arise when applying the rules of canonical quantization in the
presence of linear, dispersive and absorbing materials. Within this framework,
which we use as the basis for our numerical approach below, the medium-supported
electromagnetic field is formally generated by local bosonic dynamical operators
$\hat{\mathbf{f}}(r,\omega)$ at every point in space and frequency, with the EM
field obtained through a convolution of the EM Green's function. While the sheer
number of formal modes prevents their direct use in a ``standard'' description,
a large number of relevant observables and effects can be obtained in approaches
where these degrees of freedom are integrated out in some sense, with final
expressions only depending on the EM Green's function after performing a
perturbation expansion or treating few-level emitters approximately as bosonic
degrees of freedom~\cite{Dung1998,Wubs2004,Buhmann2012I,Buhmann2012II,
Delga2014}.

Even for the case nonperturbative interactions between several emitters and
arbitrary photonic structures, it was realized by Buhmann \emph{et
al.}~\cite{Buhmann2008Casimir}, and later independently by several other
groups~\cite{Hummer2013,Rousseaux2016}, that a unitary frequency-dependent basis
transformation can be used to transform the local operators
$\hat{\mathbf{f}}(r,\omega)$ to a set of new modes in such a way that only a
\emph{single} photonic mode interacts with each emitter at each frequency, with
the the strength of the interaction encoded in the spectral density,
$J(\mathbf{r},\omega)$ at the position $\mathbf{r}$ of the emitter. We note that
one naturally arrives at the same picture by calculating the local density of EM
states and using its relation with the decay rate and the dyadic Green's
function~\cite{Novotny2012}.

When the spectral density has a Lorentzian profile, the dynamics can be mapped
to the dissipative Rabi model~\cite{Grynberg2010,
Gonzalez-Tudela2014}. Generalizing this idea, if the spectral density is well
approximated as a sum of $N$ Lorentzians, the dynamics can be fully solved by
including $N$ dissipative bosonic modes~\cite{Delga2014,Li2016Transformation,
Cuartero-Gonzalez2018}. However, for arbitrary complex spectral densities, this
approximation is not useful. In that case, one approach is to exploit the tools
developed for open quantum systems~\cite{Carmichael1999,Gardiner2004,
DeVega2017}, which exactly describe a quantum system coupled to a continuous
``bath'' described by a given spectral density. In particular, if the coupling
between the system and the bath is weak, one can apply the Markov approximation
(which assumes that the bath has ``no memory''), such that the EM environment
simply introduces a frequency-dependent decay rate (corresponding exactly to the
Purcell effect). When this approximation is not applicable, more advanced
numerical approaches such as tensor network calculations~\cite{Schollwock2011,
Schroder2019} or hierarchical equations of motion~\cite{Tanimura1990} can be
employed, possibly after a chain transformation of the associated
Hamiltonian~\cite{Chin2010}. Such approaches have been used to study static
properties and dynamics in organic polaritons~\cite{DelPino2018Ground,
DelPino2018Dynamics}. However, these are numerically demanding approaches that
require significant computational resources.

In this work, we explore an intermediate approach that goes beyond a mean-field
description, without trying to obtain a full quantum description of the coupled
emitter-photon system. We do so by employing the cumulant or cluster expansion
method~\cite{Kubo1962,Kira2008,Kira2011} to treat the interaction of a single
quantum emitter with an arbitrary photonic spectral density. This method has its
roots in the Bogoliubov--Born--Green--Kirkwood--Yvon hierarchy
(BBGKY)~\cite{Kira2011}. It relies on the fact that for a system of interacting
particles, the dynamics of the mean value of an $N$-particle operator depend on
the mean values of $N+1$-particle operators. Truncating this description by
neglecting operator correlations above some order leads to a closed set of
equations. This method was already applied in the context of cavity
QED~\cite{Henschel2010,Kirton2018, Hoffmann2019Benchmarking, Zens2019}, but a systematic
study of the importance of the different terms appearing in the expansion has
not been provided yet. We here present an extensive study of how different
truncations of the cumulant expansion perform in the computation of the dynamics
of the quantum emitter and EM modes. In particular, we investigate the effect of
truncating the cumulant expansion at different orders and compare different
strategies for performing these truncations. To benchmark our method, we choose
spectral densities for which (almost) exact solutions can be obtained through
the Wigner-Weisskopf and dissipative Rabi model, respectively.

\section{Method}
Within the framework of macroscopic QED, the Hamiltonian that describes the
interaction between one emitter and a medium-assisted electromagnetic field is,
within the dipole approximation~\cite{Buhmann2008Casimir} (here and in the
following, we use units where $\hbar=1$),
\begin{multline}\label{Genhamiltonian}
H = \sum_{\lambda=e,m}\int d\mathbf{r}^3\int_0^{\infty} d\omega\ \omega\ \mathbf{f}_{\lambda}^{\dagger}(\mathbf{r},\omega)\mathbf{f}_{\lambda}(\mathbf{r},\omega)\\
+ H_\mathrm{em} - \hat{\bm{\mu}} \cdot \mathbf{E}(\mathbf{r}_A),
\end{multline}
where $f_{\lambda}(\mathbf{r},\omega)$ and
$f^{\dagger}_{\lambda}(\mathbf{r},\omega)$ are the bosonic annihilation and
creation operators, $H_\mathrm{em}$ is the bare-emitter Hamiltonian,
$\hat{\bm{\mu}}$ is the dipole operator of the two-level system, and
$E(\mathbf{r}_A)$ is the electric field operator, which is given by a
superposition of the bosonic operators
$\hat{\mathbf{f}}_{\lambda}(\mathbf{r},\omega)$ with weights determined by the
classical Green's tensor $\mathbf{G}(\mathbf{r}_A,\mathbf{r},\omega)$. As
mentioned above, a frequency-dependent unitary transformation of the
$f_{\lambda}(\mathbf{r},\omega)$ can be performed such that for each frequency,
only a \emph{single} photonic mode $a(\omega)$ interacts with the
emitter~\cite{Buhmann2008Casimir} (under the assumption that only a single
polarization direction interacts with the emitter dipole operator). Furthermore,
we here approximate the quantum emitter as a two-level system described by the
Pauli matrices $\sigma^i$ ($i\in\{x,y,z\}$), with transition frequency $\Omega_0$
and transition dipole moment $\bm{\mu}$. The Hamiltonian then becomes
\begin{multline}\label{MQEDhamiltonian}
H = \int_0^{\infty} \mathrm{d}\omega\ \omega a^{\dagger}(\omega)a(\omega) + \frac{\Omega_0}{2} \sigma^z + \\
\int_0^{\infty} \mathrm{d}\omega\ g(\omega)\left(a^{\dagger}(\omega) +a(\omega)\right) \sigma^x,
\end{multline}
where $g(\omega)$ is the coupling between the emitter and the electromagnetic modes,
\begin{equation}\label{continter}
g(\omega)=\sqrt{\frac{\mu_0}{\pi}\omega^2 \bm{\mu}\cdot  \text{Im}\mathbf{G}(\mathbf{r}_A,\mathbf{r}_A,\omega)\cdot \bm{\mu}},
\end{equation}
where $\mathbf{r}_A$ is the position of emitter. The expression inside the
square root in \autoref{continter} is the spectral density
$J(\mathbf{r},\omega)$. For the numerical implementation, we discretize the
frequency integrals on a grid with regular spacing $\Delta\omega$. Formally,
we define the discrete orthonormal modes
\begin{equation}
    a_n = \frac{1}{\sqrt{\Delta\omega}} \int_{n\Delta\omega}^{(n+1)\Delta\omega} a(\omega) \mathrm{d}\omega,
\end{equation}
which obey $[a_n,a_m^{\dagger}]=\delta_{nm}$ since the original continuum modes
obey $[a(\omega),a^\dagger(\omega')] = \delta(\omega-\omega')$. This leads to the
discrete Hamiltonian
\begin{equation}
  H_d = \sum_n \omega_n a^{\dagger}_n a_n +
  \frac{\Omega_0}{2} \sigma^z + \sum_n g_n \left(a^{\dagger}_n + a_n\right) \sigma^x,
\end{equation}
where $\omega_n = (n+\frac12)\Delta\omega$ and $g_n =
\sqrt{J(\mathbf{r}_A,\omega_n)\Delta\omega}$. Here, we have discarded the
(infinite number of) superpositions of $a(\omega)$ orthogonal to $a_n$ in each
interval that would make the transformation unitary. Formally, this
discretization can be understood as a chain transformation~\cite{Chin2010,Chin2011} of
the continuum modes within each interval $[n\Delta\omega,(n+1)\Delta\omega]$
under the approximation that $g(\omega)$ is constant within it, and discarding
all but the first chain site.

In order to describe the action of an incoming classical electromagnetical field
(e.g., a laser pulse), it would be possible to simply use a product of coherent
states as the initial wave function, $|\psi(0)\rangle =
\prod_n|\alpha_n(0)\rangle = \prod_n e^{\alpha_n(0) a_n^\dagger - \alpha_n(0)^*
a_n}|0\rangle$, where the $\alpha_n(0)$ correspond to the classical amplitudes
of the modes when expressing the laser pulse in the basis defined by these
modes. In order to avoid the necessity for explicitly propagating this classical
field within the quantum calculation, the classical and the quantum field can be
split in the Hamiltonian using a time-dependent displacement
operator~\cite{Cohen-Tannoudji1987} $T(t) = e^{\sum_n\alpha_n^*(t) a_n -
\alpha_n(t) a_n^\dagger}$, where $\alpha_n(t) = \alpha_n(0)e^{-i\omega_n t}$.
Applying this transformation to the wavefunction, $|\psi'\rangle =
T(t)|\psi\rangle$, corresponds to transforming the Hamiltonian as
\begin{multline}
  H'_d = T(t) H_d T^{\dagger}(t) - i T(t) \partial_{t} T^{\dagger}(t) \\
  = \sum_n \omega_n a_n^{\dagger} a_n +
  \frac{\Omega_0}{2} \sigma^z + \sum_n g_n\left(a_n^{\dagger}+a_n\right)\sigma^x \\
  - \sum_n \omega_n \alpha_n(t)\alpha_n^{*}(t) + \sum_n g_n\left(\alpha_n(t)+\alpha_n^{*}(t)\right)\sigma^x,
\end{multline}
where $\sum_n g_n\left(\alpha_n(t)+\alpha_n^{*}(t)\right)$ can be replaced by
the interaction of the classical field at the emitter position with the emitter
dipole, $-\mu \mathcal{E}(t)$, while $\sum_n \omega_n \alpha_n(t)\alpha_n^{*}(t)
= \sum_n \omega_n |\alpha_n(0)|^2$ just corresponds to a constant energy shift
that can be neglected. In the following, we thus use $\mathcal{H} = H_d -
\mu\mathcal{E}(t)\sigma^x$, i.e.,
\begin{equation}\label{hamiltonian}
  \mathcal{H} = \sum_n \omega_n a_n^{\dagger} a_n + \frac{\Omega_0}{2} \sigma^z + \sum_n g_n\left(a_n^{\dagger}+a_n\right)\sigma^x - \mu\mathcal{E}(t)\sigma^x
\end{equation}
as the effective Hamiltonian and take the initial state as the vacuum state with
the emitter in its ground state\footnote{In principle, this is not the ground
state of the full system, as we include counter-rotating terms in the
light-matter coupling, and thus ultrastrong-coupling effects such as
ground-state modifications~\cite{FriskKockum2019}. For the cases we treat below,
the error due to this approximation is negligible}. However, it is important to
remember that EM field observables are also transformed according to
\begin{equation}
\expect{\psi|O|\psi} = \expect{\psi'|T(t)OT^{\dagger}(t)|\psi'},
\end{equation}
such that, e.g., $\expect{\psi|a_n|\psi}= \expect{\psi'|a_n+\alpha_n(t)|\psi'}$.
This takes into account that the ``quantum'' field generated by the
laser-emitter interaction interferes with the classical pulse propagating
through the structure, and ensures a correct description of absorption of the
pulse, coherent scattering, and similar effects. We note that the above
properties imply that within this framework, the action of any incoming laser
pulse on the \emph{full} emitter-cavity system can be described purely by the
action of the medium-enhanced classical electric field driving the emitter, with
no additional explicit driving of any EM modes. This is in contrast to, e.g.,
input-output theory, where the EM field is split into modes inside the cavity
and free-space modes outside, and external driving thus affects the cavity
modes. It should be stressed in this context that $\mathcal{E}(t)$ is the field
obtained at the position of the emitter after propagation of the external laser
pulse through the cavity structure, i.e., it contains any field enhancement and
temporal distortion induced by the cavity. In practice, it is thus most
straightforward to employ classical EM simulations to calculate the electric
field reaching the emitter for a given input pulse and cavity structure.

\subsection{Heisenberg equations of motion}
The evolution of any expectation value $\expect{O} =
\expect{\psi'|O|\psi'}$ can be described by the Heisenberg equation of
motion
\begin{equation}\label{Heisenbergeq}
  \partial_t \expect{O} =  i\expect{[\mathcal{H},O]}.
\end{equation}
In general, the time derivative of products of $N$ operators $\expect{A_1 A_2
\dots A_N}$ includes the contribution of $N+1$ operators $\expect{A_1 A_2 \dots
A_N A_{N+1}}$ due to the bilinear matter-field coupling in
\autoref{hamiltonian}, so one obtains an infinite set of equations that describe
the system. Truncating these expansions and thus neglecting some contributions
leads to a closed set of equations. This can be done in a systematic way using
the cumulant expansion (also known as cluster expansion~\cite{Kira2008,Kira2011}
or truncated BBGKY hierarchy~\cite{Hoffmann2019Benchmarking}). The cumulant
expansion method express an expectation value as sums and products of
expectation values of a smaller number of operators and their correlations and
itself does not imply any approximation. However, it then allows to
systematically discard only high-order correlations, and not just high-order
expectation values.

As an aside, we note that the meaning of ``order of the approximation'' depends
on which set of operators is used to represent the system. For example, we use
$\sigma^x$, $\sigma^y$, and $\sigma^z$ as the ``fundamental'' operators, but it
would be equally possible to use only $\sigma^x$ and $\sigma^y$ (or more
conventionally $\sigma^\pm = \frac12\sigma^x \pm \frac i2\sigma^y$) as $\sigma^z =
\sigma^+\sigma^- - \sigma^-\sigma^+ = 2\sigma^+\sigma^- - 1$. Similarly, we only
use $a_n$ and $a_n^\dagger$, but it would be equally possible to add the number
operator $N_n$ and thus obtain photonic populations at lower orders. The
convention we use is chosen because of the direct connection to Maxwell-Bloch
and other mean-field approximations, where the population of the two-level
system is considered explicitly within the set of equations, while only the
coherent part of the EM fields is treated.

In addition to the different levels of approximation for the dynamics obtained
by truncating the systems at various orders, it should be noted that the order
of the expansion needed to describe the system also depends on the
expectation values of interest. For example, the second-order correlation
function $g^{(2)}(0)$ contains expectation values of products of four operators
and is exactly equal to unity within the mean-field approximation.

We next show the set of equations obtained in our system at various orders, and
discuss possible strategies for truncation. Some of these equations have been
obtained by using the QuantumAlgebra.jl package~\cite{QuantumAlgebra.jl} for
symbolic calculation of quantum operator expressions.

The set of equations that arise from applying \autoref{Heisenbergeq} to single
operators (i.e., at first order) are
\begin{subequations}\label{MFeq}
\begin{align}
  \label{an1}
   \partial_t \expect{a_n} &= -i\omega_n \expect{a_n} -i g_n \expect{\sigma^x},\\
   \label{sx}
   \partial_t \expect{\sigma^x} &= -i \Omega_0 \expect{\sigma^y},\\
   \label{sy}
   \partial_t \expect{\sigma^y} &= \Omega_0 \expect{\sigma^x} - 2 \sum_n g_n \expect{a_n^{\dagger}\sigma^z} + 2\mu E(t)\expect{\sigma^z},\\
   \label{sz}
   \partial_t \expect{\sigma^z} &= 2 \sum_n g_n \expect{a_n^{\dagger}\sigma^y} - 2\mu E(t)\expect{\sigma^y}.
\end{align}
\end{subequations}

Within the cumulant expansion, the expectation value of a product of operators
is expressed as $\expect{ab} = \expect{a} \expect{b} + \corr{ab}$, where
$\corr{ab}$ is the correlation between $a$ and $b$. The mean-field
approximation consists in already neglecting all two-operator correlations,
i.e., to assume $\corr{a b} \simeq 0$. If this approximation is made,
Eqs.~(\ref{MFeq}) form a closed set that can be propagated in time.

At the next order of approximation, correlations up to second order are taken
into account. The Heisenberg equations of motion that arise are then
\begin{widetext}
\begin{subequations}\label{2ordereq}
\begin{align}
  \partial_t \expect{a_n^{\dagger} \sigma^x} &=
   i\omega_n \expect{a^{\dagger}_n \sigma^x} - \Omega_0 \expect{a^{\dagger}_n \sigma^y} +i g_n,
  \label{adsx}\\
  \partial_t \expect{a^{\dagger}_n \sigma^y} &= i \omega_n \expect{a_n^{\dagger}\sigma^y} +\Omega_0 \expect{a_n^{\dagger}\sigma^x} - g_n \expect{\sigma^z} -2 \sum_m g_m \left( \expect{a_n^{\dagger} a_m^{\dagger}\sigma^z} + \expect{a_n^{\dagger} a_m\sigma^z} \right) + 2\mu \mathcal{E}(t)\expect{a_n^{\dagger} \sigma^z},
  \label{adsy}\\
  \partial_t \expect{a^{\dagger}_n \sigma^z} &= i \omega_n \expect{a_n^{\dagger}\sigma^z} + g_n\expect{\sigma^y}
  +2 \sum_m g_m \left( \expect{a_n^{\dagger} a_m^{\dagger}\sigma^y} + \expect{a_n^{\dagger}a_m \sigma^y} \right)- 2\mu \mathcal{E} (t)\expect{a_n^{\dagger}\sigma^y},
  \label{adsz}\\
  \partial_t \expect{a_n^{\dagger}a_m} &= i\left(\omega_n - \omega_m\right)\expect{a^{\dagger}_n a_m}
  + i g_n \expect{a_m\sigma^x}-i g_m \expect{a_n^{\dagger}\sigma^x},
  \label{ada}\\
  \partial_t \expect{a_n^{\dagger}a^{\dagger}_m} &= i\left(\omega_n + \omega_m\right)\expect{a^{\dagger}_n a^{\dagger}_m}
  + i g_n \expect{a^{\dagger}_m\sigma^x}+i g_m \expect{a_n^{\dagger}\sigma^x}.
  \label{add}
\end{align}
\end{subequations}
Since $\expect{a_n \sigma^i} = \expect{a_n^{\dagger}\sigma^i}^*$ and
$\expect{a_n^{\dagger}a^{\dagger}_m} = \expect{a_m a_n}^*$,
Eqs.~(\ref{2ordereq}) are enough to describe all combinations of two operators.
In the cumulant expansion, we reexpress $\expect{abc} =
\expect{a}\expect{b}\expect{c} + \expect{a}\corr{bc} + \expect{b}\corr{ac} +
\expect{c}\corr{ab} + \corr{abc}$. For completeness, we here give the equations
of motion of the correlations explicitly:
\begin{subequations}
\label{CE-2ordereq}
\begin{align}
  \partial_t \corr{a_n^{\dagger} \sigma^x} &=
   i\omega_n \corr{a^{\dagger}_n \sigma^x} - \Omega_0 \corr{a^{\dagger}_n \sigma^y} +i g_n\left( 1- \expect{\sigma^x} \expect{\sigma^x} \right),
  \label{cadsx}\\
  \begin{split}
    \partial_t \corr{a^{\dagger}_n \sigma^y} &= i \omega_n \corr{a_n^{\dagger}\sigma^y} +\Omega_0 \corr{a_n^{\dagger}\sigma^x} - g_n\left( \expect{\sigma^z} - i\expect{\sigma^x}\expect{\sigma^y} \right) + 2\mu \mathcal{E}(t)\corr{a_n^{\dagger} \sigma^z}\\
    &-2 \sum_m g_m \left( \expect{a_m^{\dagger}} \corr{a_n^{\dagger}\sigma^z} + \expect{a_m} \corr{a_n^{\dagger}\sigma^z} + \expect{\sigma^z} \corr{a_n^{\dagger}a^{\dagger}_m} +
    \expect{\sigma^z} \corr{a_n^{\dagger}a_m} + \corr{a_n^{\dagger}a_m^{\dagger}\sigma^z} +
    \corr{a_n^{\dagger}a_m\sigma^z} \right),
    \label{cadsy}
  \end{split}\\
  \begin{split}
    \partial_t \corr{a^{\dagger}_n \sigma^z} &= i \omega_n \corr{a_n^{\dagger}\sigma^z} + g_n\left( \expect{\sigma^y} - i\expect{\sigma^x} \expect{\sigma^z} \right) - 2\mu \mathcal{E}(t)\corr{a_n^{\dagger} \sigma^y}\\
    & +2 \sum_m g_m \left( \expect{a_m^{\dagger}} \corr{a_n^{\dagger}\sigma^y} + \expect{a_m} \corr{a_n^{\dagger}\sigma^y} + \expect{\sigma^y} \corr{a_n^{\dagger}a^{\dagger}_m} + \expect{\sigma^y} \corr{a_n^{\dagger}a_m} + \corr{a_n^{\dagger}a_m^{\dagger}\sigma^y} + \corr{a_n^{\dagger}a_m\sigma^y} \right),
    \label{cadsz}
  \end{split}\\
  \partial_t \corr{a_n^{\dagger}a_m} &= i\left(\omega_n - \omega_m\right)\corr{a^{\dagger}_n a_m}
  + i g_n \corr{a_m\sigma^x}-i g_m \corr{a_n^{\dagger}\sigma^x},
  \label{cada}\\
  \partial_t \corr{a_n^{\dagger}a^{\dagger}_m} &= i\left(\omega_n + \omega_m\right)\corr{a^{\dagger}_n a^{\dagger}_m}
  + i g_n \corr{a^{\dagger}_m\sigma^x}+i g_m \corr{a_n^{\dagger}\sigma^x}.
  \label{cadd}
\end{align}
\end{subequations}
\end{widetext}

The cumulant expansion provides a systematic approach to approximate the true
solution by neglecting higher-order correlations between operators. A priori,
one could assume that it is always a better approximation to neglect a
correlation $\corr{A_1\ldots A_n}$ than the corresponding expectation value
$\expect{A_1\ldots A_n}$ directly. However, as we will see later, this
assumption is not always correct and whether to neglect correlations or
expectation values is a better approximation depends on the physical system and
concrete situation.

We also mention that while the equations~(\ref{2ordereq}) describing the
expectation values are linear, the corresponding correlation expansion,
Eqs.~(\ref{CE-2ordereq}) corresponds to a nonlinear system depending on products
of the state variables (expectation values and correlations). These
nonlinearities make the obtained set of equations numerically more unstable.

In equations~(\ref{CE-2ordereq}), no approximations have been made, as no
correlations have been neglected yet. To obtain a closed set of equations that
may allow the description of the time evolution of the system, some correlations
have to be neglected again. The second-order cumulant expansion approximation
means to neglect the correlations of three of more operators, ($\corr{a b
c}\simeq 0$), so that the set of equations~(\ref{MFeq}) and (\ref{CE-2ordereq})
are enough to find a solution. The same procedure as above can be followed to
obtain the equations up to third order, i.e., neglecting correlations of four or
more operators (for reference, the required cumulant expansion is given in
Appendix~\ref{app:cumulants}). The third-order expectation values needed to describe
the third order completely are $\expect{a_n^{\dagger}a_m\sigma^{x,y,z}}$,
$\expect{a_n^{\dagger}a_m^{\dagger}\sigma^{x,y,z}}$,
$\expect{a_n^{\dagger}a_m^{\dagger}a_l}$ and
$\expect{a_n^{\dagger}a_m^{\dagger}a_l^{\dagger}}$, with their explicit
equations of motion given in Appendix~\ref{appA:3rdorder}. The equations for the
correlations are not written, but it is straightforward to derive them from the
equations of motion of the expectation values.

The numerical implementation of the equations is performed within the Julia
programming language~\cite{Bezanson2017}. The code runs on graphical processing
units (GPUs), which provides a significant speedup ($\approx20$ for our
available setup) over the CPU variant of the same code. For the time
propagation, we rely on the DifferentialEquations.jl
package~\cite{Rackauckas2017}.

\section{Results}
\subsection{Free space dynamics}
The spectral density of an emitter in free space is
\begin{equation}
    J(\omega)=\hbar \omega^3 \frac{\left|\mu\right|^2}{6\pi \epsilon_0 c^3}.
\end{equation}
Discretizing this spectral density with frequency spacing $\Delta\omega$ is
equivalent to describing an emitter in center of a spherical box of radius
$R=\pi c/\Delta\omega$~\cite{Steck2007}, where $c$ is the speed of light in
vacuum. As a first example, we will treat spontaneous emission from an initially
excited emitter, i.e., the classical Wigner-Weisskopf
problem~\cite{Weisskopf1930}. If time propagation is performed over too long
times ($\approx 2R/c = 2\pi/\Delta\omega$, the time that it takes the photon to
propagate from the emitter to the boundary of the sphere and back), artificial
reflections of the emitted photons from the boundaries of the sphere are
obtained and interact again with the emitter. As the lifetime of typical
emitters (atoms, molecules, quantum dots) is on the scale of nanoseconds, an
accurate description would require a very small frequency spacing and thus a
very large box, and additionally, propagation over very long times. To avoid
this, we instead set the emitter dipole moment to the unrealistically large
value of $\mu = 2565$~D, for which the spontaneous emission lifetime at the
emitter frequency of $\Omega_0=2.72$~eV is given by $\tau \approx 46$~fs. We
choose $N=400$ photonic modes on a regular grid in frequency from $0$~eV to
$5.44$~eV. For these parameters, spontaneous emission takes place within a time
shorter than $2R/c$.

\begin{figure}[t]
  \includegraphics[width=\linewidth]{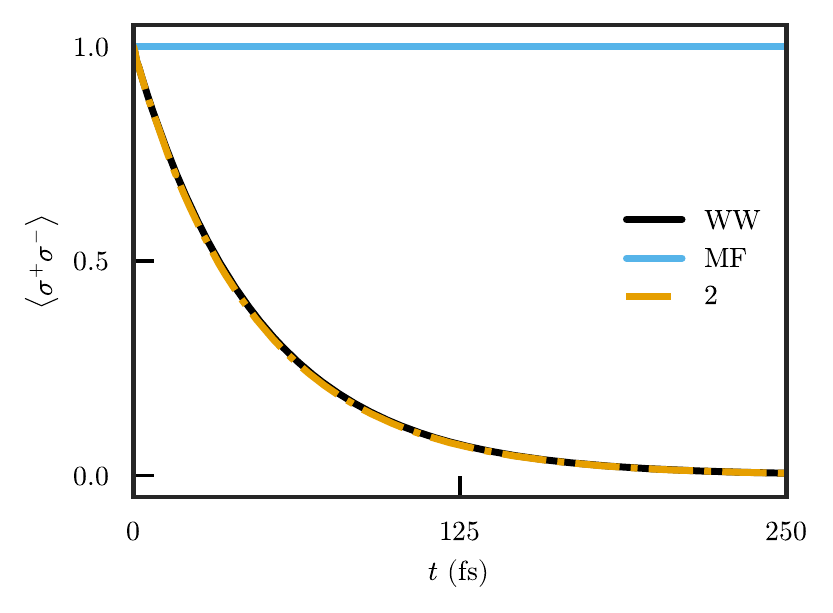}
  \caption{\label{SEfree} Excited-state population of the emitter in time in
  free space. Comparison between Wigner-Weisskopf approximation (silver line),
  mean-field (light green line), and second-order approximation (medium green
  dash-dotted line).}
\end{figure}

The spontaneous emission dynamics of an initially excited single emitter in free
space is shown in \autoref{SEfree}, which shows the excited-state population of
the emitter as a function of time, calculated using three different numerical
methods: Perturbative Wigner-Weisskopf theory (WW), which simply predicts
exponential decay with rate $\gamma = J(\Omega_0)$, mean-field (MF), and
second-order cumulant expansion (2). As is well-known, the mean-field
approximation does not predict any spontaneous emission. This is because this
phenomenon is due to the interaction of the emitter with the vacuum fluctuations
$\expect{a_n^{\dagger} a_m}$ and the mean-field approximation neglects all the
expectation values of two or more operators. Since $\expect{a_n} =
\expect{\sigma^x} = \expect{\sigma^y} = 0$ at $t=0$ and no external electric
field affects the system, no dynamics are predicted. Going beyond mean-field is
thus essential to describe spontaneous emission~\cite{Andreasen2009,Chen2019I}.
On the other hand, the second-order cumulant expansion (and all higher-order
approaches, not shown) already perfectly describes the free-space spontaneous
decay of $\expect{\sigma^{+}\sigma^{-}}$ due to the vacuum fluctuations.

\begin{figure}[t]
  \includegraphics[width=\linewidth]{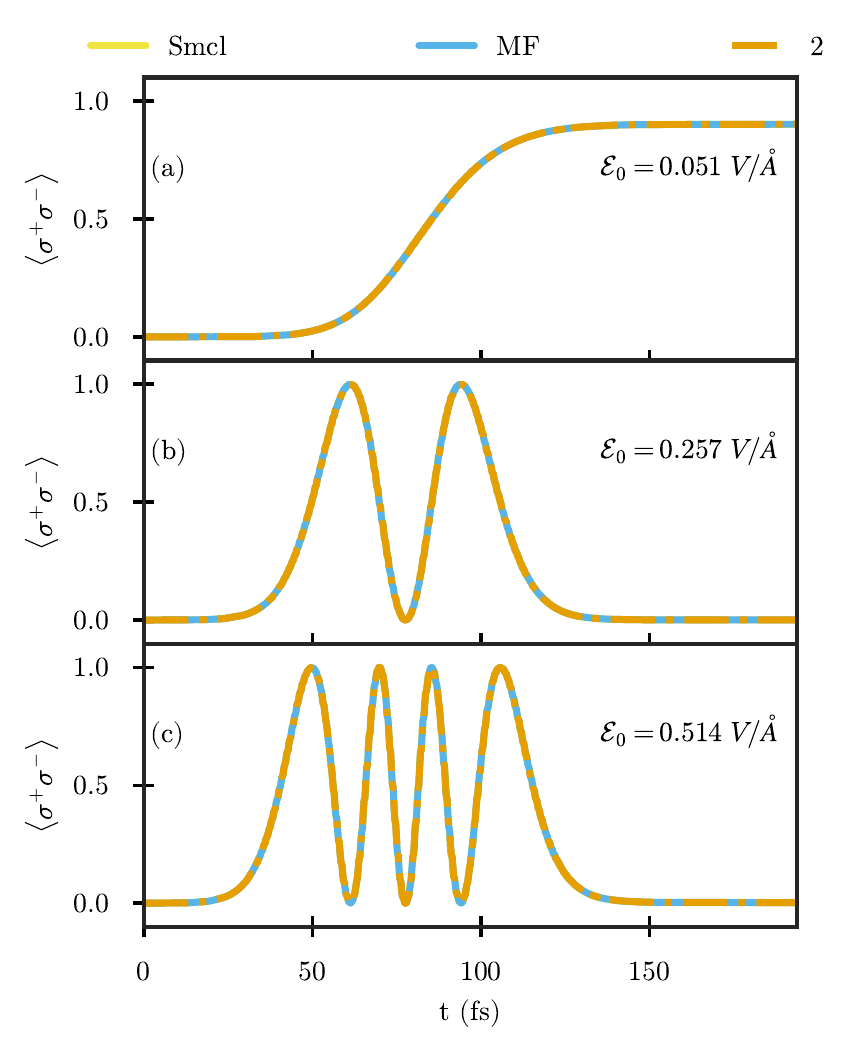}
  \caption{Excited-state population of a single emitter in free space under
  driving by a short Gaussian pulse. The peak electric field amplitude increases
  from (a) to (c) and is shown in each subplot. Comparison between
  semi-classical approximation (yellow line), mean-field (light green line) and
  second-order approximation (dot-dashed medium green line). All lines are
  indistinguishable for this case.}
  \label{SPfree}
\end{figure}

We next consider an emitter initially in its ground state,
$\expect{\sigma^{+}\sigma^{-}}(t=0) = 0$ under pumping by a classical electric
field $\mathcal{E}(t)$. We use a short Gaussian pulse in resonance with the
emitter transition frequency, $\mathcal{E}(t)=\mathcal{E}_0
e^{-(t-t_0)^2/2T^2}\sin\left(\Omega_0 t\right)$. In order to describe a more
realistic system, the dipole moment of the emitter is set to $\mu=2.56$~D,
corresponding to a spontaneous emission lifetime in free space of $\tau\approx
46$~ns. The pulse parameters are $t_0=77.76$~fs and $T=24.20$~fs. We now compare
the mean-field and second-order approaches with a semi-classical approximation
in which no quantized light modes are present at all, and the two-level system
interacts with the EM field via the equations
\begin{equation}
    \frac{\partial}{\partial t}\begin{pmatrix} C_g\\ C_e \end{pmatrix} = -i \begin{pmatrix}
        0 & -\mu\cdot \mathcal{E}(t)\\
       -\mu \cdot\mathcal{E}(t)  & \Omega_0
    \end{pmatrix} \cdot \begin{pmatrix} C_g\\ C_e \end{pmatrix},
\end{equation}
where $C_g$ and $C_e$ are the ground-state and excited-state amplitudes,
respectively. The peak amplitudes of the electric field we consider are
$\mathcal{E}_0=0.051$~V/\r{A} (\autoref{SPfree}a),
$\mathcal{E}_0=0.257$~V/\r{A} (\autoref{SPfree}b) and
$\mathcal{E}_0=0.514$~V/\r{A} (\autoref{SPfree}c). For the
weakest driving we consider, the system is already in the nonlinear regime but the 
electric field is weak enough so that no Rabi oscillations are seen in the atom dynamics
(subplot a), while the two stronger fields lead to a strongly nonlinear response
with driven Rabi oscillations (subplots b and c). In this case, the coupling to
the free-space modes is so weak that they are not expected to have any influence
on the dynamics, and this is indeed observed in \autoref{SPfree}. All three
approaches (semi-classical, mean-field, and second order) accurately describe
the emitter dynamics, and correlations between the photonic modes and the
emitter can be neglected. After the end of the pulse, the spontaneous decay
(with lifetime $\tau\approx 46$~ns) is so slow that it is not noticeable over
the timescales we investigate, although it would show up eventually for longer
propagation times for the second-order approach.

\subsection{Cavity}
We next consider a spectral density that represents a single lossy cavity mode.
This is achieved using a Lorentzian frequency dependence,
\begin{equation}
    J(\omega) = \frac{g^2}{\pi}\frac{\gamma/2}{\left(\omega-\omega_c\right)^2 + \left(\gamma/2\right)^2}.
\end{equation}
The dynamics predicted using this spectral density is mathematically equivalent
to those of the Lindblad master equation
\begin{equation}
\partial_t \rho = -i\left[H_{R},\rho\right] + \gamma \mathcal{L}_a[\rho]
\end{equation}
where $H_{R}$ is the Rabi Hamiltonian
\begin{equation}
    H_{R}= \frac{\Omega_0}{2}\sigma^z + \omega_c a^{\dagger}a + g(a +
    a^{\dagger}) \sigma^x
\end{equation}
for interaction of a single emitter with a single quantized mode, while
$\mathcal{L}_a[\rho]=a\rho a^\dagger -\frac12 \{a^\dagger a,\rho\}$ is the
Lindblad operator that describes the cavity losses. The effective coupling $g =
\mu E_{1ph}$ is determined by the amplitude of the Lorentzian spectral density,
the effective losses are given by its width $\gamma$, and the frequency of the
photonic mode is the resonance frequency $\omega_c$ of the
Lorentzian~\cite{Grynberg2010, Gonzalez-Tudela2014}.

The emitter and cavity frequencies are both set to $\Omega_0=\omega_c=2.72$~eV.
We choose a bandwidth of $\gamma=0.027$~eV (Q-factor $Q=\omega_c/\gamma = 100$)
and will consider various coupling amplitudes $g$. The number of modes
considered is $N=400$ and the frequencies are taken from $\omega_{\min}=2.04$~eV
to $\omega_{\max}=3.40$~eV, (grid spacing $\Delta\omega = 3.4$~meV), so the
range is wide enough and the number of modes big enough to represent the
Lorentzian spectral density.

\begin{figure}[t]
  \includegraphics[width=\linewidth]{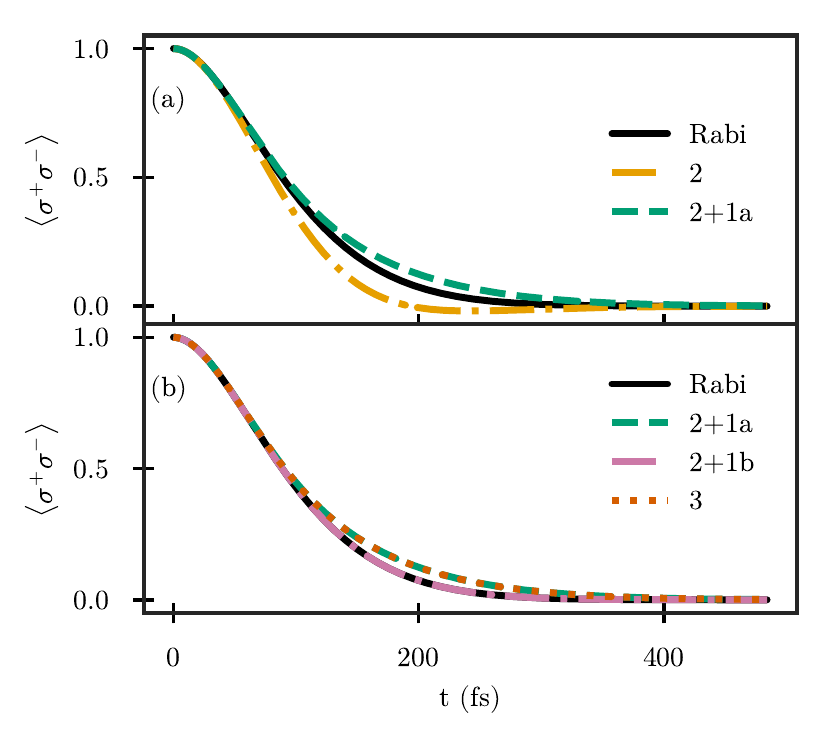}
  \caption{\label{SE-8d8} Excited-state population for an initially excited
  emitter in a cavity for coupling $g=0.008$~eV. (a) Comparison between the
  exact Rabi model solution (purple line), $2^\mathrm{nd}$ order (medium green
  line) and 2+1a (dashed dark green line) approximation. (b) Comparison between
  the Rabi solution (purple line), 2+1a approximation (dashed dark green line),
  2+1b approximation (dashed dotted blue line), and $3^\mathrm{rd}$ order
  (dotted red line).}
\end{figure}

The emitter is initially in its excited state $\expect{\sigma^+\sigma^-}(t=0)=1$
and evolves freely in the cavity, without any external electric field. While we
do not employ the rotating wave approximation (RWA), which consists in
neglecting the counter-rotating terms $a_n^\dagger \sigma^+$ and $a_n \sigma^-$
in the Hamiltonian, it is approximately fulfilled for the coupling values we
choose here. Within the RWA, the number of excitations $\sigma^+\sigma^- +
\sum_n a_n^\dagger a_n$ is conserved. In \autoref{SE-8d8}, the evolution of the
emitter population is shown for a coupling strength of $g=0.008$~eV, for which
the system is already close to the strong-coupling regime
($4g>\gamma$)~\cite{Torma2015}. In contrast to the free-space case, the
second-order approximation (shown in \autoref{SE-8d8}a) now starts to show some
differences with respect to exact solution obtained with the Rabi model, with
the population even reaching nonphysical values, $\expect{\sigma^+\sigma^-}<0$.
This implies that some third-order terms are required to obtain the correct
dynamics, but it is not clear a priori which additional terms have to be
included. We thus compare different extensions of the second-order expansion by
successively adding higher-order terms. In the first one, the second-order set
of equations (\ref{MFeq}) and (\ref{CE-2ordereq}) is used, but in
\autoref{cadsy}, the term $\corr{a_n^{\dagger}a_m\sigma^z}$ and its dynamics are
not neglected. We denote this second-order approximation with a correction by
``2+1a'' in the following. The importance of including the particular
third-order term has been previously pointed out in the
literature~\cite{Henschel2010}, and is due to it being the third-order
correction with the largest value. Taking into account that, to a good
approximation, the state during the dynamics is described by a single
excitation, $|\psi\rangle \approx (\alpha \sigma^+ + \sum_n \beta_n a_n^\dagger)
|0\rangle$, we can easily see this by inspecting the cumulant expansion of the
third-order expectation values. For $\expect{a_n^{\dagger}a_m\sigma^z}$, this
gives
\begin{multline}
    \expect{a_n^{\dagger}a_m\sigma^z} = \expect{a_n^{\dagger}} \expect{a_m} \expect{\sigma^z} + \expect{a_n^{\dagger}} \corr{a_m \sigma^z} + \\
    \expect{a_m} \corr{a_n^{\dagger}\sigma^z} + \expect{\sigma^z} \corr{a_n^{\dagger}a_m} + \corr{a_n^{\dagger}a_m\sigma^z}.
\end{multline}
The first three terms are negligible since $\expect{a_n}\approx0$, but the
product $\expect{\sigma^z} \corr{a_n^{\dagger}a_m}$ is non-negligible since both
the emitter and photonic mode populations are nonzero. At the same time, it does
not approximate the value of $\expect{a_n^{\dagger}a_m\sigma^z}$ well, so that
the correlation $\corr{a_n^{\dagger}a_m\sigma^z}$ is necessarily non-zero. In
contrast, the expansion of $\expect{a_n^{\dagger}a_m\sigma^y}$ gives
\begin{multline}
    \expect{a_n^{\dagger}a_m\sigma^y} = \expect{a_n^{\dagger}} \expect{a_m}
    \expect{\sigma^y} + \expect{a_n^{\dagger}} \corr{a_m \sigma^y} + \\
    \expect{a_m} \corr{a_n^{\dagger}\sigma^y} + \expect{\sigma^y}
    \corr{a_n^{\dagger}a_m} + \corr{a_n^{\dagger}a_m\sigma^y}.
\end{multline}
Here, all the product terms contain at least one negligible value as
$\expect{\sigma^y}\approx 0$, while $\expect{a_n^{\dagger}a_m\sigma^y}$ is also
zero for the single-excitation state given above. This implies that the
correlation $\corr{a_n^{\dagger}a_m\sigma^y}$ is in turn also negligible.

In the equation of motion of the new term $\expect{a_n^{\dagger}a_m\sigma^z}$,
fourth-order expectation values appear, see \autoref{adasz} in the appendix.
Performing the cumulant expansion on these and neglecting the fourth-order
correlation, it is easy to see that only third-order correlations that are
neglected in the other equations appear, and by consistency, these terms are
approximated up to the second order as well, leading to \autoref{2+1a}.

\begin{figure}[t]
  \includegraphics[width=\linewidth]{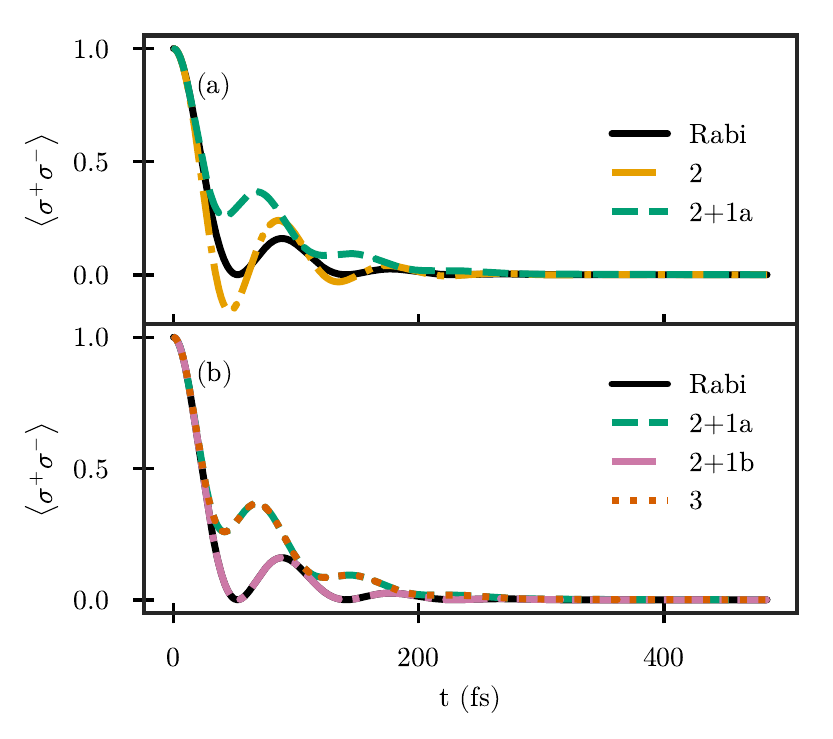}
  \caption{\label{SE-8d7}
  Excited-state population of an initially excited emitter in a cavity in the
  strong-coupling regime, with $g=0.024$~eV. Subplots and lines like in
  \autoref{SE-8d8}.}
\end{figure}

\begin{figure}[t]
  \includegraphics[width=\linewidth]{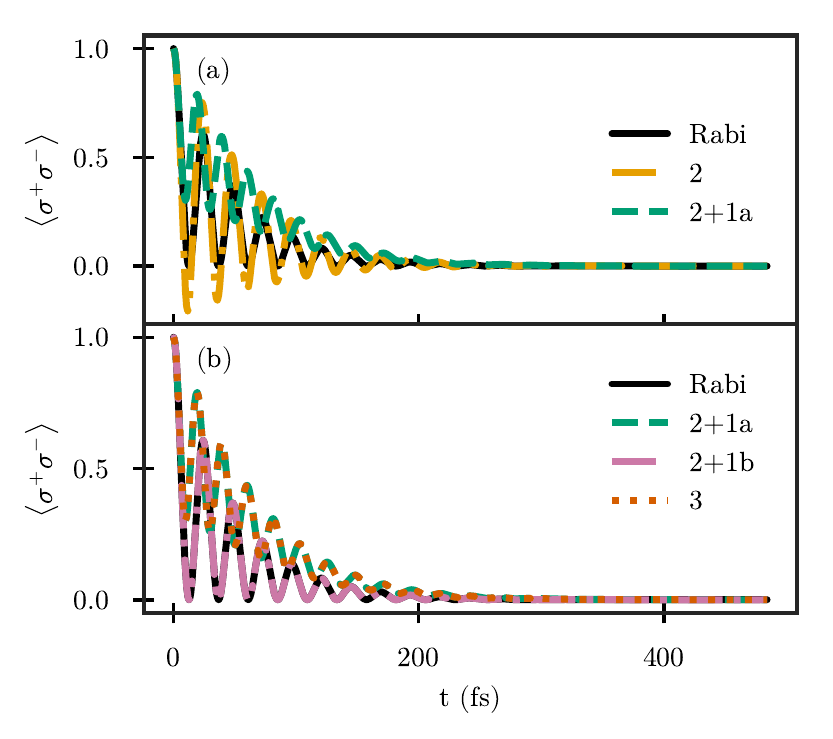}
  \caption{\label{SE-1d5} 
  Excited-state population of an initially excited emitter in a cavity in the
  strong-coupling regime, with $g=0.086$~eV. Subplots and lines like in
  \autoref{SE-8d8} and \autoref{SE-8d7}.}
\end{figure}

As seen in \autoref{SE-8d8}a, the dynamics of the emitter within the 2+1a
approximation are changed, with the population never reaching negative values.
However, it still does not agree with the exact solution provided by the Rabi
model, and is now overestimated. Inspection of the equation of motion for
$\expect{a_n^{\dagger}a_m\sigma^z}$, \autoref{adasz}, shows that this contains
fourth-order order terms that involve two photonic creation or annihilation
operators. For the current dynamics, where to a good approximation only one
excitation is present in the system, these fourth-order expectation values are
thus approximately zero. Within the 2+1a approximation, they are however
represented by products of non-negligible second-order correlations that have
non-zero values. It is then possible to improve the approximation by not
performing a cumulant expansion on the fourth-order terms in \autoref{adasz},
but by neglecting them directly. We are going to refer to this approximation as
``2+1b''. When using it (shown in \autoref{SE-8d8}b), the emission dynamics are
now correctly obtained.

Finally, we also perform the full third-order expansion, with the cumulant
expansion performed on all expectation values and fourth-order correlations
being neglected. The third-order approximation (also shown in \autoref{SE-8d8}b)
provides identical results as the 2+1a approach, proving that, indeed, all
third-order correlations apart from $\corr{a_n^{\dagger}a_m\sigma^z}$ can be
neglected. However, for good agreement with the exact results, the same
correction as in the 2+1b approach would have to be performed, or alternatively
the full expansion would have to be performed up to at least fourth order. 

We note here that numerically, both approximations 2+1a and 2+1b are only
slightly more costly than the second-order expansion, since the added term
$\corr{a_n^{\dagger}a_m\sigma^z}$ only contains two continuum indices $n$ and
$m$. In contrast, the full third-order expansion contains terms of the form
$\corr{a_n a_m a_o}$ with three continuum mode indices (represented by $N\times
N\times N$ arrays), and is thus significantly more expensive to implement.

\begin{figure}[t]
  \includegraphics[width=\linewidth]{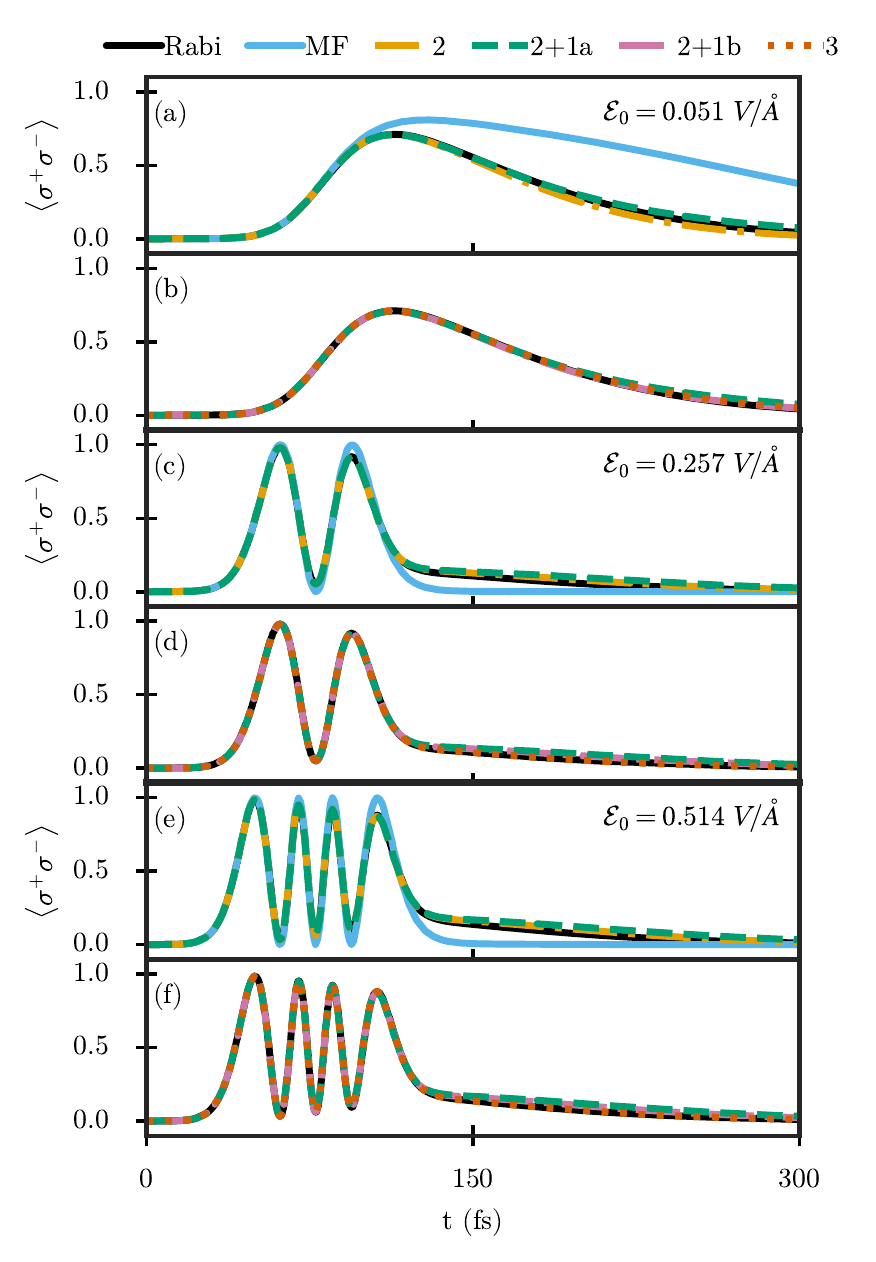}
  \caption{\label{Sshort8d8}
  Excited-state population for an emitter initially in the ground state within a
  cavity with $g=0.008$~eV, driven by a short classical electric field pulse
  (see text for details), for three different peak amplitudes as indicated in
  the subplots. (a), (c), (e) Comparison between the Rabi solution (purple
  line), mean-field (light green line), $2^\mathrm{nd}$ order (medium green
  dash-dotted line) and 2+1a approximation (dark green dashed line). (b), (d),
  (f) Comparison between Rabi solution (purple line), 2+1a approach (dashed dark
  green line), 2+1b approach (dashed dotted blue line), and $3^\mathrm{rd}$
  order (dotted red line).}
\end{figure}

\begin{figure}[t]
  \includegraphics[width=\linewidth]{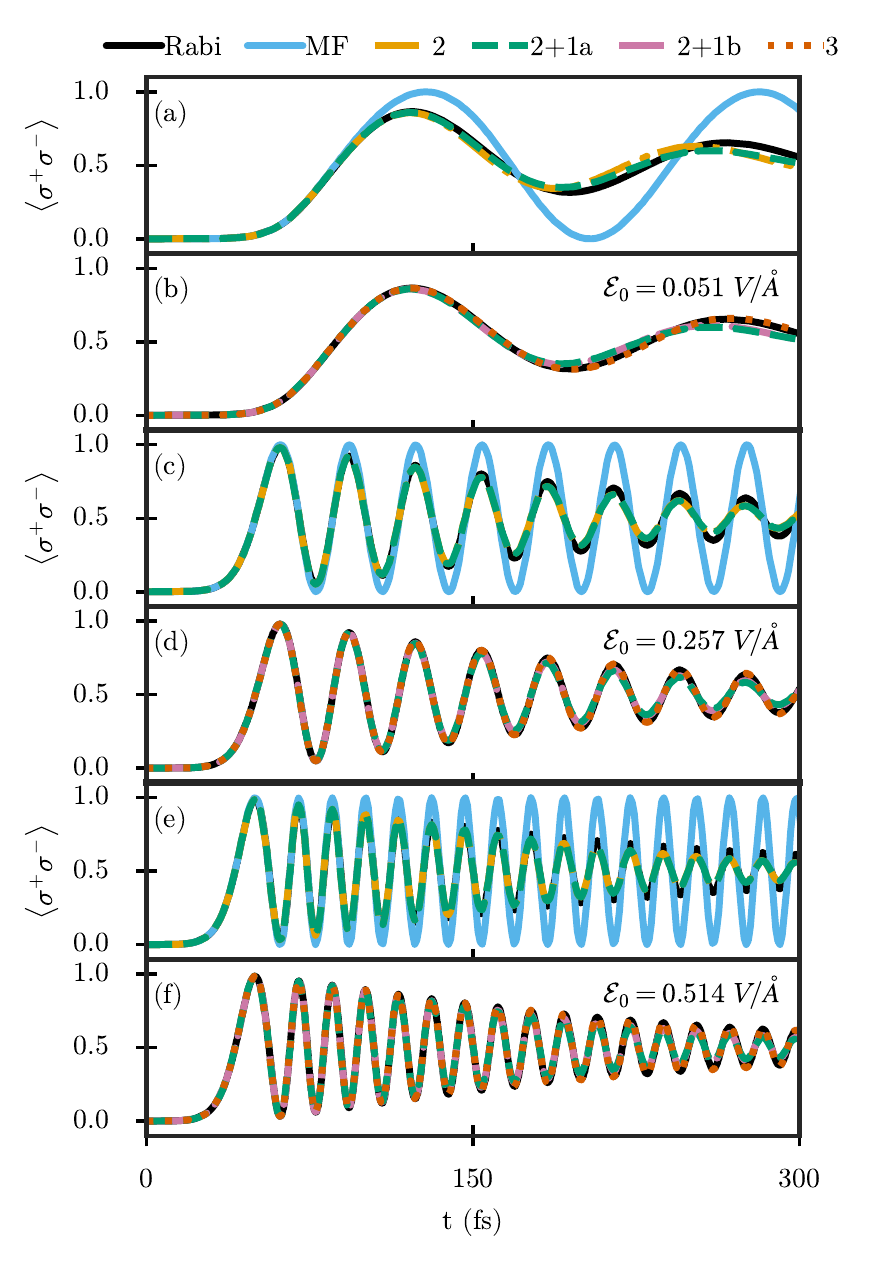}
  \caption{\label{Slong8d8} 
  Like \autoref{Sshort8d8}, but for a semi-infinite driving pulse with constant
  amplitude after a smooth turn-on (see text for details).}
\end{figure}

We now increase the coupling strength to $g=0.024$~eV, squarely in the
strong-coupling regime where vacuum Rabi oscillations are expected, and study
the emitter dynamics as shown in \autoref{SE-8d7}. All the previous
approximations are compared again. The nonphysical values that the population
takes in the second-order approximation are more evident when the coupling
increases, although the Rabi oscillation frequency is reproduced well. For the
2+1a approximation, which again gives identical results as the full third-order
expansion, this does not hold. This is because the correlations that appear in
the equation of motion of $\corr{a_n^{\dagger}a_m\sigma^z}$ interact via the
coupling, so if the coupling increases, the modifications produced by the
spurious correlations also increase. The correction 2+1b, i.e., enforcing the
fourth-order expectation value in \autoref{adasz} to be zero, again predicts the
exact dynamics accurately since the system remains in the single-excitation
subspace even in strong coupling. These conclusions are essentially unchanged
even when increasing the coupling to $g=0.086$~eV, shown in \autoref{SE-1d5}.

We next compare the same physical system and the same approximations, but now
not for the case of spontaneous emission and vacuum Rabi oscillations, but for
the emitter initially in its ground state, $\expect{\sigma^+ \sigma^-}(t=0)=0$,
and driven by an incoming classical electric field. Two different pulses are
considered. First, we take the same short Gaussian pulse considered in free
space $\mathcal{E}(t)=\mathcal{E}_0 e^{-(t-t_0)^2/2 T^2}\sin(\Omega_0 t)$
(assumed to be the pulse reaching the emitter after enhancement and distortion
by propagating through the cavity structure). In the second case, we choose an
electric field that smoothly turns on and then remains at a stationary intensity
indefinitely, $\mathcal{E}(t)= \mathcal{E}_0 \sin(\Omega_0 t)
(\theta(t_0-t)e^{-\left(t-t_0\right)^2/2T^2} + \theta(t-t_0))$, where
$\theta(t)$ is the Heaviside theta function, allowing to study if and how a
steady state is reached in the time propagation. In both cases, the pump laser
frequency is in resonance with the emitter and cavity resonances.

We first again use the cavity with the weakest light-matter coupling
($g=0.008$~eV). The emitter population dynamics when a Gaussian pulse excites
the system is shown in \autoref{Sshort8d8}, while the steady-state pulse is
shown in \autoref{Slong8d8}. In both figures we compare the same approximations
as above with the exact Rabi model solution. The amplitudes of the electric
field interacting with the emitter are the same as in free space, given by
$\mu\mathcal{E}_0=0.026$~eV (subplots a and b),
$\mu\mathcal{E}_0=0.132$~eV (subplots c and d), and
$\mu\mathcal{E}_0=0.263$~eV (subplots e and f).

In contrast to the free-space case, the quantum effects due to fluctuations,
such as spontaneous emission, are not negligible here, and the mean-field
approximation (shown in subplots a, c and e) fails to capture the dynamics as it
can only represent the coherent contribution to the light-matter
interaction~\cite{Chen2019I}. In the short-pulse case, \autoref{Sshort8d8}, this
is mostly seen in the dynamics after the pulse, but is also reflected in Rabi
oscillations during the pulses with bigger amplitudes than the ones predicted by
the exact solution. Still, the mean-field approximation does give a
qualitatively correct prediction of the behavior for the short-pulse case,
\autoref{Sshort8d8}. In the case of a long pulse, \autoref{Slong8d8}, the
initial driven oscillations are well-described but, as there is no coupling
between the fluctuations and the emitter, no steady state is achieved and the
population keeps oscillating indefinitely. If the decay rate of the emitter is
known, incoherent contributions to the emitter dynamics can be incorporated
\emph{ad hoc} using phenomenological decay constants~\cite{Allen1987}. However,
obtaining these constants is not always easy and is only straightforward in the
weak-coupling regime where the light and matter degrees of freedom are not
mixed. In those case, the validity and simplicity of the mean-field
approximation makes it a common tool in describing a wide range of systems
pumped by lasers~\cite{Cuerda2016,Grynberg2010}.

The second-order approximation (subplots a, c and e) is sufficient to describe
the dynamics in this regime. When the amplitude of the electric field is
$\mathcal{E}_0=0.051$~V/\r{A} and the pulse is short (\autoref{Sshort8d8}a), the
dynamics predicted by this approximation are much more similar to the Rabi
solution, as incoherent contributions are taken into account via the
second-order terms. For a long pulse (\autoref{Slong8d8}a), the oscillations are
not accurately described, neither in shape nor in amplitude, but it does give a
qualitative prediction and the steady state is predicted quantitatively. Making
the correction 2+1a to the second order changes the dynamics only slightly. The
extra correlations included by this correction lead to a decrease of the
oscillation amplitude, but the qualitative description is maintained. Finally,
enforcing the fourth order expectation values to be zero via the correction 2+1b
(subplot b), i.e, enforcing the system to have only one excitation, hardly
changes the prediction of the emitter dynamics.

If the driving electric field is more intense (subplots c, d, e and f) the
second-order approximation (subplots c and e) gives a correct description of the
shape of the Rabi oscillations, but their amplitude is underestimated.
Approximations 2+1a and 2+1b (subplots d and f) do not show any difference with
respect to the ``bare'' second order. Thus, correlations that change the
description of the dynamics completely in the case of spontaneous emission do
not matter much in the more classical case of driving by a strong laser pulse.
Finally, the third-order approximation is shown in subplots b, d and f. Adding
all the third-order correlations sufficiently modifies the dynamics to achieve
an accurate prediction in good agreement with the Rabi model. 

From the results in \autoref{Sshort8d8} and \autoref{Slong8d8}, we can conclude
that when the light-matter coupling is not too strong, the second-order
correlations are the most important and in general this order of approximation
is enough to describe the main characteristics of the solution. If a more
quantitative description is required, the third-order approximation achieves
almost perfect agreement with the exact dynamics.

\begin{figure}[t]
  \includegraphics[width=\linewidth]{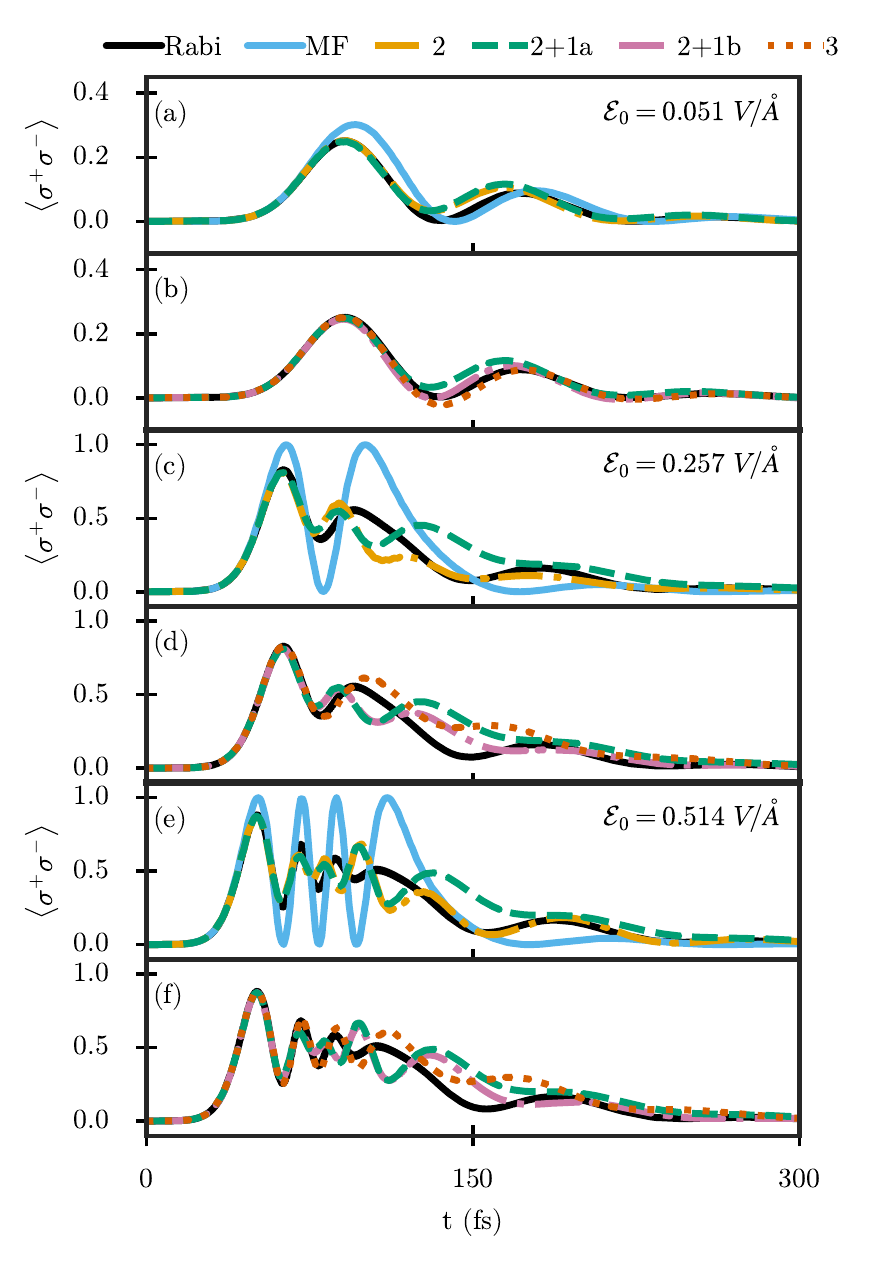}
  \caption{\label{Sshort8d7}
  Like \autoref{Sshort8d8}, but for emitter-cavity coupling of $g=0.024$~eV.}
\end{figure}

\begin{figure}[t]
  \includegraphics[width=\linewidth]{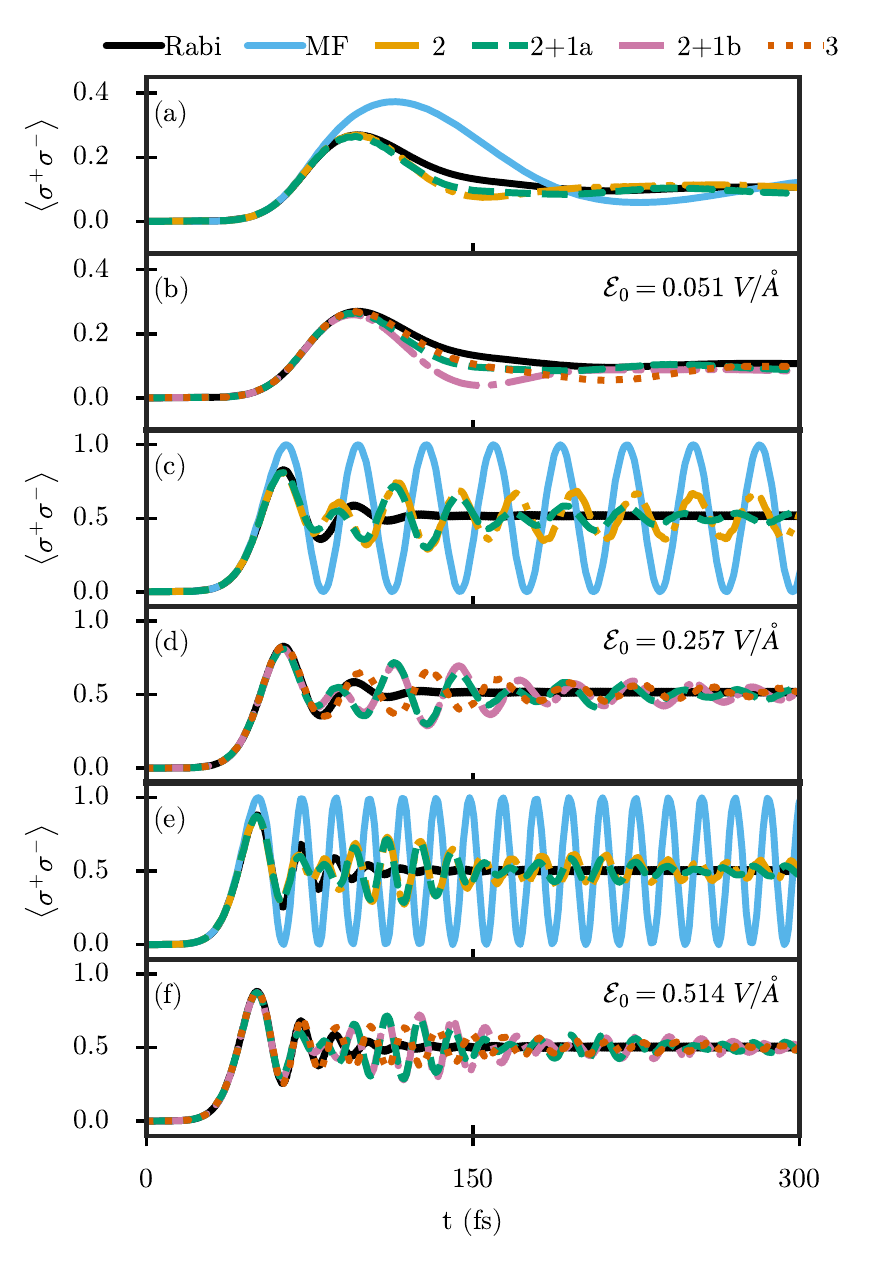}
  \caption{\label{Slong8d7}
  Like \autoref{Sshort8d7}, but for a semi-infinite driving pulse with constant
  amplitude after a smooth turn-on.}
\end{figure}

The results obtained when again increasing the light-matter coupling strength to
$g=0.024$~eV are shown in \autoref{Sshort8d7} for the Gaussian pulse and in
\autoref{Slong8d7} for the semi-infinite pulse, with the same driving pulses as
in the previous case. When the amplitude of the electric field is
$\mathcal{E}_0=0.051$~V/\r{A} (subplots a and b), its magnitude is comparable to
the coupling strength. The mean-field approximation (subplot a) then
overestimates the population oscillations for both classical fields. This
continues for more intense driving fields (subplots c and e). As mentioned
above, while the mean-field approximation cannot reproduce spontaneous decay by
itself, adding phenomenological decay constants to the mean-field equations can
be used to achieve reasonable descriptions of the strong-coupling regime for
intense classical fields~\cite{Cuerda2015}. However, doing so means that the
photons emitted due to field fluctuations are not represented, so that, e.g.,
the spontaneous emission from polaritonic states~\cite{Coles2014} could not be
monitored in the emitted field.

Compared to the mean-field approach, the second-order approximation better
predicts both the short-time dynamics as well as the steady-state limit for the
semi-infinite pulse for the weak driving amplitude $\mathcal{E}_0 =
0.051$~V/\r{A}, but slightly overestimates the population at intermediate times.
This overall picture also applies for the stronger driving strengths (subplots
c-f). The corrections 2+1a and 2+1b somewhat improve upon the bare second-order
calculation, with 2+1a working slightly better for the semi-infinite pulses,
\autoref{Slong8d7}, and 2+1b working slightly better under short-pulse driving,
\autoref{Sshort8d7}. Finally, as could be expected, the third-order
approximation improves the results for both the short and semi-infinite pulses.
In particular, it perfectly reproduces the exact results during the first few
Rabi oscillations, and converges to the correct steady-state limit under
long-pulse driving faster than the lower-order expansions. However, even the
third-order expansion does not fully reproduce the dynamics at intermediate
times, where decoherence starts to set in and induces corrections to the
coherent dynamics, which are reflected in higher-order light-matter correlations
at intermediate times. At longer times, where the system becomes mostly
incoherent, the light-matter correlations are again well-described by
lower-order expansions, and the steady state is thus well-represented within the
third-order and even second-order expansions.

\begin{figure}[t]
  \includegraphics[width=\linewidth]{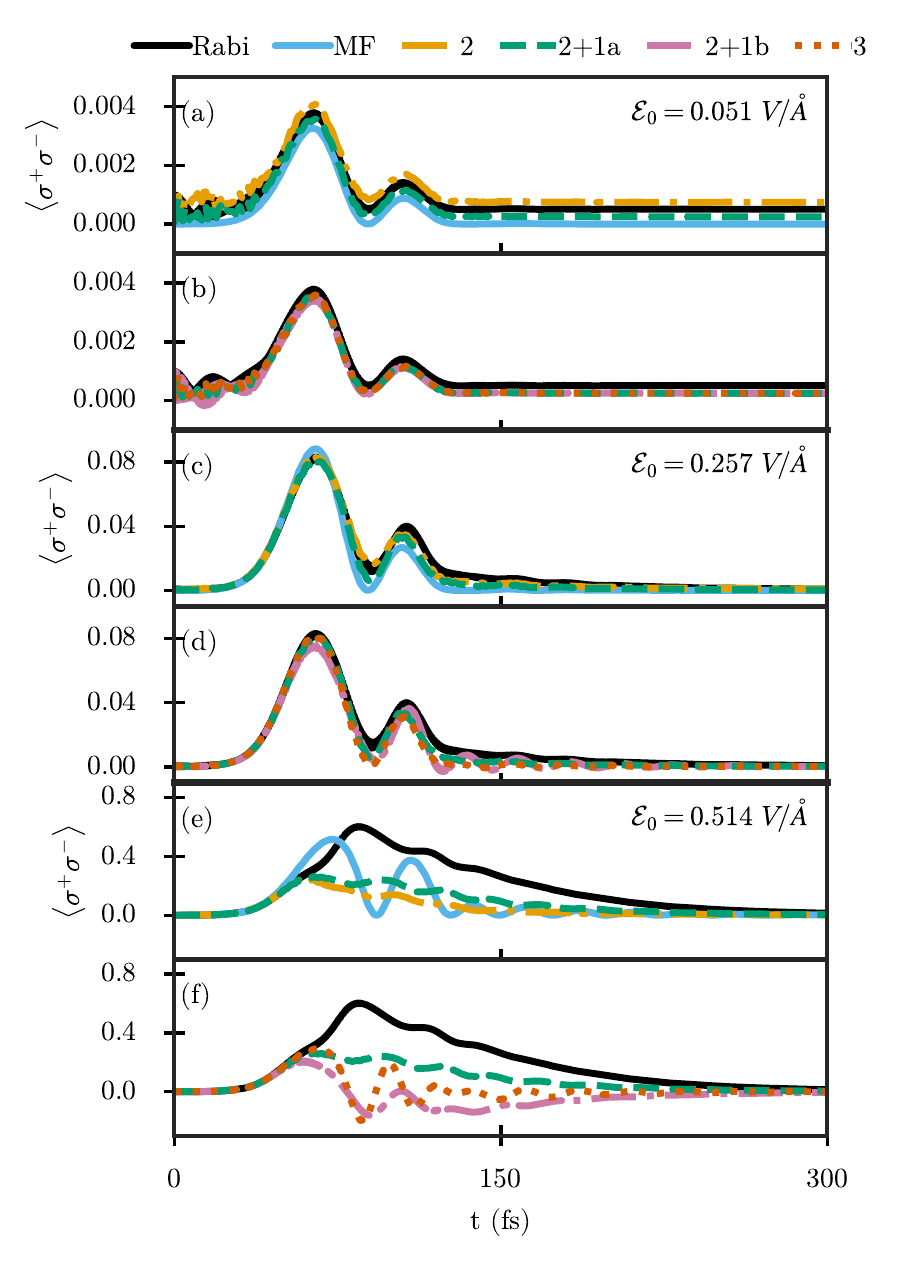}
  \caption{\label{Sshort1d5}
  Like \autoref{Sshort8d8}, but for emitter-cavity coupling of $g=0.086$~eV.}
\end{figure}

\begin{figure}[t]
  \includegraphics[width=\linewidth]{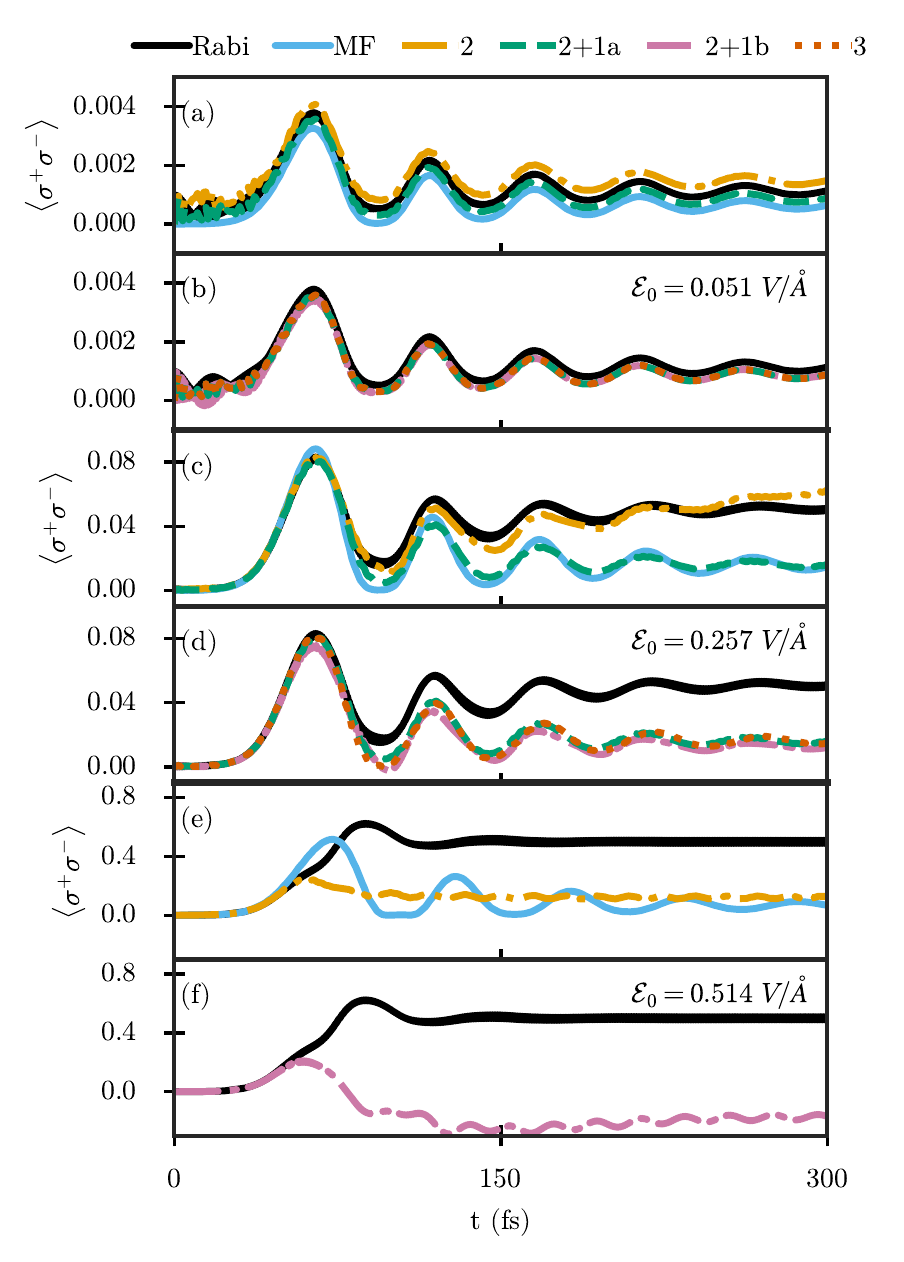}
  \caption{\label{Slong1d5}
  Like \autoref{Sshort1d5}, but for a semi-infinite driving pulse with constant
  amplitude after a smooth turn-on.}
\end{figure}

To push the approximations more to their limit, we now increase the
emitter-cavity coupling to $g=0.086$~eV and again show the emitter population
dynamics under short-pulse driving, in \autoref{Sshort1d5}, and for a
semi-infinite pulse, in \autoref{Slong1d5}. The amplitudes of the classical
electric fields and their parameters are the same as in the previous figures.
For these parameters, we are approaching the ultrastrong-coupling
regime~\cite{FriskKockum2019}, as the Rabi splitting $\Omega_R \approx 2g =
0.172$~eV becomes non-negligible compared to the emitter frequency $\Omega_0 =
2.72$~eV. This implies that the counter-rotating terms in the light-matter
interaction become important and even the ground state becomes dressed. Although
we still choose the uncoupled ground state of the system (EM vacuum and emitter
in the ground state) as the initial state, this state is not the ground state of
the coupled system, and starting the dynamics immediately leads to fast
``quenching'' or ``ringdown'' oscillations at short times. These are seen for
weak driving fields in subplots a and b of \autoref{Sshort1d5} and
\autoref{Slong1d5}. Additionally, the very strong coupling implies that the
polaritonic states of the coupled cavity-emitter system at $\omega_{\pm} \approx
\Omega_0 \pm g$ are now quite strongly detuned from the driving pulse that is
tuned to resonance with the bare-emitter (and cavity) resonance frequency. The
excitation amplitudes and driven Rabi oscillation frequencies in this case are
therefore significantly smaller than for the previously treated systems with
smaller light-matter coupling strengths.

We now again investigate the validity of the various approximations. The
mean-field approximation cannot represent the ultrastrong-coupling induced
changes, which only show up in correlations but do not lead to coherent fields.
Therefore, neither the ground state nor the steady state of the system can  be
described correctly. This is especially noticeable under weak driving (subplots
a and b), where the shape of the driven oscillations is predicted reasonably
well, but the final populations are underestimated for both types of driving.
The higher-order expansions improve on this result, but not even the third-order
approximation manages to fully reproduce the dynamics. This failure is most
likely due to the fact that the low-order correlation expansions now have to
reproduce both the ultrastrong-coupling induced correlations as well as the
driving-pulse induced correlations, so that overall, higher-order correlations
become more important than in cases with weaker emitter-cavity coupling. Still,
under weak driving, all approximations manage to represent the overall dynamics
reasonably well up to a global shift. Interestingly, in this case, the 2+1a,
2+1b and third-order approximations all perform almost identically.

When the driving field amplitude is increased (subplots c and d in
\autoref{Sshort1d5} and \autoref{Slong1d5}), the correction to the population
due to the counter-rotating terms becomes less noticeable since the
laser-induced populations are larger. However, the predictions of the cumulant
expansion methods start to diverge more and more from the exact results obtained
within the dissipative Rabi model. Here, the (ultra)strong light-matter coupling
in combination with the strong driving induces large correlations between light
and matter that fail to be described within low-order cumulant expansions. In
particular, in the case of the semi-infinite pulse, the results obtained within
the cumulant expansion fail to reproduce the steady-state results even
qualitatively and lead to significant shifts. It should be noted that these
effects are expected to be less relevant when many emitters are included in the
cavity~\cite{Kirton2017}.

For the most intense driving field (subplots e and f in \autoref{Sshort1d5} and
\autoref{Slong1d5}), all considered approximations start to break down for the
strong emitter-cavity coupling considered here. For the short-pulse case,
\autoref{Slong1d5}, none of the approximations reproduces the Rabi model even
qualitatively, with the 2+1b and third-order results again reaching unphysical
values of the emitter population, $\expect{\sigma^+\sigma^-}<0$.

For the case of the semi-infinite pulse, \autoref{Slong1d5}, a similar picture
presents itself. For these strong driving pulses, none of the approximations
captures the emitter dynamics well. In particular, the simulations using the
2+1a and third-order approximations break down even more dramatically shortly
after the start of the pulse, with the emitter population diverging towards
infinity. These results are therefore not shown here. We note that, as far as we
could determine, these divergences are not due to numerical issues that could be
solved by using better integration algorithms, but correspond to the actual
behavior of the system description at the chosen level, and thus indicate a
complete breakdown of the approximations.

\section{Summary and Outlook}
To summarize, we have explored the cumulant expansion method to calculate the
Heisenberg equations of motion for one emitter coupled to an arbitrary number of
EM modes with an arbitrary spectral density, as obtained through the formalism
of macroscopic QED in nanophotonic and plasmonic systems. In order to benchmark
the method, we have compared its results to two well-known cases where
quasi-exact solutions are available: An emitter in free-space, where
perturbative approaches to light-matter coupling are valid, and a Lorentzian
spectral density that can be mapped analytically to a Lindblad master equation
describing the dissipative Rabi model, i.e., coupling of the emitter to a single
cavity mode with losses. In the case of the cavity, we have explored the change
in behavior as the coupling strength is increased from the weak up to the
ultrastrong-coupling regime. We have investigated both the spontaneous emission
dynamics where the emitter is initially excited and the behavior when a
classical pulse pumps the system and compared exact solutions with the
predictions at different orders of approximation. We have found that, in order
to describe spontaneous emission, going beyond the mean-field is essential.
While in free space, the second-order approximation is enough to describe this,
in the cavity the fact that the photon can be reabsorbed after emission leads to
corrections that are only well-described at higher orders of approximation.
Here, we have identified a single third-order term that describes the only
important contribution at that order, $\corr{a_n^\dagger a_m \sigma^z}$. In
order to describe spontaneous emission correctly (using the approximation we
call 2+1b), it is then necessary to explicitly disregard a fourth-order
expectation value, instead of performing the cumulant expansion on it. More
systematic approximations, such as 2+1a, in which no specific assumptions are
made for any the expectation values of the system, cannot describe the
spontaneous emission unless higher orders are included in the expansion, as some
non-negligible correlations arise in the set of equations. For this situation,
the correlation expansion does not actually provide a better approximation than working
directly with expectation values and discarding higher orders.

As expected, the mean-field approximation is able to describe the emitter
dynamics when a classical field pumps the system if coherent interactions are
predominant. In free space, the description is accurate, although the slow
(nanosecond-scale) spontaneous emission and associated decay after the pulse
again cannot be represented. The second-order approximation again can reproduce
this decay.

In the strong-coupling regime, i.e., when the emitter is coupled to a cavity
mode with coupling strengths similar to or larger than the cavity losses, the
second-order approximation fails to describe the dynamics in several cases. The
combined action of the coherent driving laser pulse and the strong light-matter
coupling with the cavity mode lead to an increase of light-matter correlations
at intermediate times which is proportional to both the driving field strength
and the light-matter coupling strength. In order to describe these correlations
well, the order of the expansion has to be increased, with the third-order
expansion being sufficient to describe most investigated cases. At later times,
either after the pulse in short-pulse driving, or when a steady state is
approached under continuous driving, the required order of the approximation
needed to describe the system well again decreases. However, for large enough
emitter-cavity coupling strengths and driving intensities, the cumulant
expansions at the orders used here fail to describe the dynamics and become
unstable. In general, the order of approximation or even the validity of the
cumulant expansion method to describe the emitter dynamics depends strongly on
the physical system and the initial conditions and driving.

Going forward, it would be interesting to study the convergence properties of
the cumulant expansion when the number of emitters is increased. In that case,
the system is expected to behave more ``classically'' so that low-order cumulant
expansions could provide a better approximation than in the cases studied here,
in particular under driving by external coherent laser pulses. Furthermore, the
capability of the method to treat an arbitrary spectral density could be
exploited to study emitter dynamics in systems that are not well-described by a
single or few cavity modes, such as found in complex nanoplasmonic or hybrid
plasmonic-dielectric structures~\cite{Chikkaraddy2016,Li2016Transformation,
Rousseaux2016,Gurlek2018,Franke2019}.

\begin{acknowledgments}
This work has been funded by the European Research Council through grant
ERC-2016-StG-714870, and by the Spanish Ministry for Science, Innovation, and
Universities -- Agencia Estatal de Investigación through grants
RTI2018-099737-B-I00, PCI2018-093145 (through the QuantERA program of the
European Commission), and MDM-2014-0377 (through the María de Maeztu program for
Units of Excellence in R\&D).\\[0.1cm]
\end{acknowledgments}

\appendix
\section{Cumulant expansions up to fourth order}\label{app:cumulants}
For reference, we here give the cumulant expansion for expectation values of
products of up to four operators expressed in terms of single-operator
expectation values and cumulants\cite{Kubo1962}.
\begin{subequations}
  \begin{align}
    &\expect{ab} = \corr{ab} + \expect{a}\expect{b}\\
    \begin{split}
      &\expect{abc} = \corr{abc} + \expect{a}\expect{b}\expect{c} +\\
      &\qquad \expect{a}\corr{bc} + \expect{b}\corr{ac} + \expect{c}\corr{ab}
    \end{split}\\
    \begin{split}
      &\expect{abcd} = \corr{abcd} + \expect{a} \expect{b} \expect{c} \expect{d} +\\
      &\quad \expect{a} \expect{b} \corr{cd} + \corr{ab} \expect{c} \expect{d} + \corr{ab} \corr{cd} +\\
      &\quad \expect{a} \expect{c} \corr{bd} + \corr{ac} \expect{b} \expect{d} + \corr{ac} \corr{bd} +\\
      &\quad \expect{a} \expect{d} \corr{bc} + \corr{ad} \expect{b} \expect{c} + \corr{ad} \corr{bc} +\\
      &\quad \expect{a} \corr{bcd} + \expect{b} \corr{acd} + \expect{c} \corr{abd} + \expect{d} \corr{abc}.
    \end{split}
\end{align}
\end{subequations}

\section{Third-order equations}\label{appA:3rdorder}
For reference, we here reproduce the equations needed to describe the
third-order expectation values:
\begin{widetext}
\begin{subequations}
\begin{align}
\partial_t\expect{a_n^{\dagger}a_m\sigma^x} &= i\left(\omega_n - \omega_m\right)\expect{a_n^{\dagger}a_m\sigma^x} -\Omega_0 \expect{a_n^{\dagger}a_m\sigma^y} + i g_n \expect{a_m} -i g_m \expect{a_n^{\dagger}},
\label{adasx}\\
\begin{split}
  \partial_t\expect{a_n^{\dagger}a_m\sigma^y} &= i\left(\omega_n - \omega_m\right)\expect{a_n^{\dagger}a_m\sigma^y} +\Omega_0 \expect{a_n^{\dagger}a_m\sigma^x} - g_n \expect{a_m\sigma^z} - g_m \expect{a_n^{\dagger}\sigma^z}\\
  &-2\sum_l g_l \left( \expect{a_n^{\dagger}a^{\dagger}_l a_m\sigma^z}  + \expect{a_n^{\dagger}a_l a_m\sigma^z} \right) + 2 \mu \mathcal{E}(t)\expect{a_n^{\dagger}a_m \sigma^z},
\label{adasy}
\end{split}\\
\begin{split}
  \partial_t\expect{a_n^{\dagger}a_m\sigma^z} &= i\left(\omega_n - \omega_m\right)\expect{a_n^{\dagger}a_m\sigma^z} + g_n \expect{a_m\sigma^y} + g_m \expect{a_n^{\dagger}\sigma^y}\\
  &+2\sum_l g_l \left( \expect{a_n^{\dagger}a^{\dagger}_l a_m\sigma^y}  + \expect{a_n^{\dagger}a_l a_m\sigma^y} \right) - 2 \mu \mathcal{E}(t)\expect{a_n^{\dagger}a_m \sigma^y},
  \label{adasz}
\end{split}\\
\partial_t\expect{a_n^{\dagger}a^{\dagger}_m\sigma^x} &= i\left(\omega_n + \omega_m\right)\expect{a_n^{\dagger}a^{\dagger}_m\sigma^x} -\Omega_0 \expect{a_n^{\dagger}a^{\dagger}_m\sigma^y} + i g_n \expect{a^{\dagger}_m} +i g_m \expect{a_n^{\dagger}},
\label{addsx}\\
\begin{split}
  \partial_t\expect{a_n^{\dagger}a^{\dagger}_m\sigma^y} &= i\left(\omega_n + \omega_m\right)\expect{a_n^{\dagger}a^{\dagger}_m\sigma^y} +\Omega_0 \expect{a_n^{\dagger}a^{\dagger}_m\sigma^x} - g_n \expect{a^{\dagger}_m\sigma^z} - g_m \expect{a_n^{\dagger}\sigma^z}\\
  &-2\sum_l g_l \left( \expect{a_n^{\dagger}a^{\dagger}_l a^{\dagger}_m\sigma^z}  + \expect{a_n^{\dagger}a^{\dagger}_m a_l \sigma^z} \right) + 2 \mu \mathcal{E}(t)\expect{a_n^{\dagger}a^{\dagger}_m \sigma^z},
  \label{addsy}
\end{split}\\
\begin{split}
  \partial_t\expect{a_n^{\dagger}a^{\dagger}_m\sigma^z} &= i\left(\omega_n + \omega_m\right)\expect{a_n^{\dagger}a^{\dagger}_m\sigma^z} + g_n \expect{a^{\dagger}_m\sigma^y} + g_m \expect{a_n^{\dagger}\sigma^y}\\
  &+2\sum_l g_l \left( \expect{a_n^{\dagger}a^{\dagger}_l a^{\dagger}_m\sigma^y}  + \expect{a_n^{\dagger}a^{\dagger}_m a_l \sigma^y} \right) - 2 \mu \mathcal{E}(t)\expect{a_n^{\dagger} a^{\dagger}_m \sigma^y},
  \label{addsz}
\end{split}\\
\partial_t\expect{a_n^{\dagger}a^{\dagger}_m a_l} &= i\left(\omega_n + \omega_m -\omega_l\right)\expect{a_n^{\dagger}a^{\dagger}_m a_l} + i g_n \expect{a^{\dagger}_m a_l \sigma^x} +i g_m \expect{a_n^{\dagger} a_l \sigma^x} -i g_l \expect{a_n^{\dagger}a_m^{\dagger}\sigma^x},
\label{adda}\\
\partial_t\expect{a_n^{\dagger}a^{\dagger}_m a^{\dagger}_l} &= i\left(\omega_n + \omega_m +\omega_l\right)\expect{a_n^{\dagger}a^{\dagger}_m a_l} + i g_n \expect{a^{\dagger}_m a^{\dagger}_l \sigma^x} +i g_m \expect{a_n^{\dagger} a^{\dagger}_l \sigma^x} +i g_l \expect{a_n^{\dagger}a_m^{\dagger}\sigma^x}.
\label{addd}
\end{align}
\end{subequations}
 In the approximation 2+1a, just the equation (\ref{adasz}) is added to the sets of equations (\ref{MFeq}) and (\ref{2ordereq}). Moreover, the fourth-order terms are expanded up to second order, so both the fourth and the third order correlations are neglected. The equation of motion of the third-order correlation is
\begin{multline}
\partial_t \corr{a_n^{\dagger}a_m\sigma^z} = g_n\left(\corr{a_m\sigma^y} -i\expect{\sigma^x} \corr{a_m\sigma^z} -i\expect{\sigma^z} \corr{a_m\sigma^x}\right)+ g_m\left(\corr{a_n^{\dagger}\sigma^y} +i\expect{\sigma^x} \corr{a_n^{\dagger}\sigma^z} +i\expect{\sigma^z} \corr{a_n^{\dagger}\sigma^x}\right) +\\
i\left(\omega_n-\omega_m\right)\corr{a_n^{\dagger}a_m\sigma^z}+ 2\sum_l g_l\left(\corr{a_l^{\dagger}a_n^{\dagger}} \corr{a_m\sigma^y} + \corr{a_l^{\dagger}a_m}\corr{a_n^{\dagger}\sigma^y} + \corr{a_n^{\dagger}a_l} \corr{a_m\sigma^y} + \corr{a_l a_m} \corr{a_n^{\dagger}\sigma^y}\right),
\label{2+1a}
\end{multline}
\end{widetext}
In \autoref{2+1a}, the terms $\corr{a_n^{\dagger}a_m a_l\sigma^y}$, $\corr{a_m
a_l \sigma^y}$, $\corr{a_n^{\dagger}a_l\sigma^y}$ and
$\corr{a_n^{\dagger}a_m\sigma^y}$ are neglected. The correlations inside the
last bracket in (\ref{2+1a}) make this term non-negligible, although the
expectation values in (\ref{adasz}) are analytically zero.

In the approximation 2+1b, the equation (\ref{adasz}) is again the only one added but instead of doing the cumulant expansion of the higher-order terms that appear in this equation, the condition $\expect{a_n^{\dagger}a_l^{\dagger}a_m\sigma^y} = \expect{a_n^{\dagger}a_l a_m\sigma^y}=0$ is imposed directly.

\bibliography{references}

\end{document}